\documentclass[jcp,aip,reprint,superscriptaddress,showpacs,floats,floatfix,twocolumns]{revtex4-1}
\pdfoutput=1
\usepackage[english]{babel}
\usepackage{color}
\usepackage{graphicx}
\usepackage{placeins}
\usepackage[caption=false]{subfig} 
\usepackage{amsmath,amssymb,bm}
\usepackage[version=3]{mhchem}
\usepackage{verbatim}
\usepackage{multirow}
\usepackage{dcolumn}
\usepackage{float}
\usepackage{nicefrac}
\usepackage{siunitx}
\usepackage{tikz}
\usepackage{pgfplots}
\usepackage{shellesc}
\usepackage{array}
\usepackage{booktabs}

\usepackage[colorlinks,allcolors=black,citecolor=blue,urlcolor=blue]{hyperref}

\begin{document}

\title{%
Quantum Nature of the Hydrogen Bond from Ambient 
Conditions 
down to Ultra--low Temperatures
}

\author{Christoph Schran}
\email{Christoph.Schran@rub.de}
\affiliation{Lehrstuhl f\"ur Theoretische Chemie,
  Ruhr--Universit\"at Bochum, 44780 Bochum, Germany}
\author{Dominik Marx}
\affiliation{Lehrstuhl f\"ur Theoretische Chemie,
  Ruhr--Universit\"at Bochum, 44780 Bochum, Germany}
\date{\today}


\begin{abstract}
    Many experimental techniques
such as tagging photodissociation and helium nanodroplet isolation spectroscopy
    operate at 
    very low temperatures in order to investigate hydrogen bonding.
To elucidate the differences between such ultra--cold
and usual ambient conditions,
    different hydrogen bonded systems are studied systematically
from 300~K
    down to about 1~K using path integral
    simulations that explicitly consider both, the
    quantum nature of the nuclei 
and thermal fluctuations.
    For that purpose, finite sized water clusters, specifically
    the water dimer and hexamer, 
protonated water clusters including the Zundel and Eigen complexes, 
    as well as 
hexagonal ice 
as a condensed phase representative 
are compared directly as a function
    of temperature.
While weaker hydrogen bonds, as present in the neutral
systems, show distinct structural differences between
ambient conditions and the ultra-cold regime,
the stronger hydrogen bonds of the protonated
water clusters are less perturbed by
temperature compared to their quantum ground state.
In all studied systems,
    the quantum delocalization of the 
nuclei 
is found to
    vary drastically with temperature.
Interestingly, upon reaching temperatures of about 1~K, the spatial quantum 
delocalization of the heavy oxygens approaches that of the protons 
for relatively weak spatial constraints,
and even significantly exceeds the latter
in case of the centered hydrogen bond in the Zundel complex. 
    These findings are 
relevant for 
comparisons
between experiments on hydrogen bonding carried out at ultra-cold versus ambient conditions
as well as
to understand quantum delocalization phenomena of nuclei
by seamlessly extending our
insights into 
noncovalent 
    interactions
down to ultra-low temperatures. 
\end{abstract}

\maketitle

\section{Introduction}
\label{sec:intro}
The hydrogen bond has been the focus
of numerous experimental studies
some of which operate at rather low
temperatures.
Especially finite sized clusters are 
often
investigated by
means of tagging photodissociation~\cite{Chakrabarty2013/10.1021/jz402264n,
Wolk2014/10.1021/ar400125a,
Roithova2016/10.1021/acs.accounts.5b00489}
or helium nanodroplet isolation~\cite{Goyal1992/10.1103/PhysRevLett.69.933,
Toennies1998/10.1146/annurev.physchem.49.1.1,
Stienkemeier2001/10.1063/1.1415433,
Toennies2004/10.1002/anie.200300611,
Stienkemeier2006/10.1088/0953-4075/39/8/R01}
spectroscopy techniques
at temperatures on the order of~10 or even 1~Kelvin. 
Such spectroscopic techniques have been very successful in order
to study hydrogen bonding in
neutral water clusters~\cite{Liu1996/10.1038/381501a0,%
Cruzan1996/10.1126/science.271.5245.59,%
Schwan2016/10.1039/c6cp04333j,%
Schwan2019/10.1002/anie.201906048,%
Yang2019/10.1126/science.aaw4086%
},
proton transfer in HCl-water clusters~\cite{Gutberlet2009/10.1126/science.1171753, %
Mani2019/10.1126/sciadv.aav8179%
},
as well as
the microsolvation of the proton
by water~\cite{Okumura1986/10.1063/1.451079,%
Headrick2005/10.1126/science.1113094,%
Heine2013/10.1021/ja401359t,%
Fournier2015/10.1021/acs.jpca.5b04355,%
Wolke2016/10.1126/science.aaf8425,%
Fagiani2016/10.1039/c6cp05217g%
}
under cryogenic conditions.
But also condensed phase hydrogen bonded systems such
as hexagonal ice have been studied down to
temperatures of about 1~K for example
by neutron defraction~\cite{Fortes2018/10.1107/S2052520618002159}.
Other temperature studies of ice I$_\text{h}$ have revealed
interesting transitions in the thermal expansion coefficient around
100~K~\cite{Buckingham2018/10.1103/PhysRevLett.121.185505}
and also below 70~K, where ice I$_\text{h}$ features
negative thermal expansion~\cite{Roettger1994/10.1107/S0108768194004933,Roettger2012/10.1107/S0108768111046908}.

While such conditions 
significantly suppress 
thermal fluctuations and therefore provide usually
a clean picture of the hydrogen bond,
the system of interest is
predominantly
in
its ground state, which drastically increases
the importance of nuclear quantum effects (NQEs).
These quantum effects such as tunneling and zero point
vibrations
need to be incorporated in any realistic simulation on this topic
if quantitative or even qualitative conclusions are of interest.
However, already 
at moderately low or even ambient temperatures, the classical
description of the nuclei misses some important features,
in particular when hydrogen bonds are
involved~\cite{Marx1997/10.1126/science.275.5301.817,Benoit1998/10.1038/32609,Marx1999/10.1038/17579,Tuckerman2001/10.1103/PhysRevLett.86.4946,Tuckerman2002/10.1038/nature00797,Benoit2005/10.1002/cphc.200400533,Car2008/10.1103/PhysRevLett.101.017801,Markland2012/10.1088/0953-8984/24/28/284126,Ceriotti2013a/10.1073/pnas.1308560110,Wang2014/10.1063/1.4894287,Schran2017/10.1016/j.cplett.2017.04.034}.
In addition, estimating the influence of NQEs on hydrogen bonding is no trivial task,
since the quantum nature of the nuclei is known to show
competing effects~\cite{Manolopoulos2009/10.1063/1.3167790,Michaelides2011/10.1073/pnas.1016653108}.
On 
the
one hand, a 
weakening
of the 
hydrogen
bond is caused by the delocalization
of the 
shared proton
along the direction perpendicular to the hydrogen bond.
On the other hand, the 
covalent OH~bond gets elongated
due to anharmonicity, thus 
resulting in shorter and 
thereby
stronger hydrogen bonds.
The degree of compensation of these competing effects is in 
general hard to predict
since the 
shape of protons in hydrogen bonds are known since long to depend on the intermolecular bond length~\cite{Benoit2005/10.1002/cphc.200400533},
and offers more surprises when studied systematically~\cite{Schran2017/10.1016/j.cplett.2017.04.034}. 
Independently from these phenomena,
NQEs are starting to play a vital role in the correct description
of the system of interest
when decreasing the temperature.
At the same time, their inclusion in quantum simulations
increases the computational cost
dramatically
when reaching temperatures as low as 1~K. 
Only recently, the efficient study of hydrogen bonded systems in 
that
ultra-cold regime has become feasible via colored noise
thermosttating methods~\cite{Uhl2016/10.1063/1.4959602,Schran2018/10.1021/acs.jctc.8b00705}.
This enables
us 
to systematically
investigate temperature effects on hydrogen bonding
seamlessly from ambient conditions to 1~K
while 
explicitly including the quantum nature of the nuclei.

Hence, different hydrogen bonded systems,
in particular the water dimer and hexamer
as well as the three smallest
protonated water clusters, the Zundel cation (\cf{H5O2+}),
the protonated water trimer (\cf{H7O3+}),
and the Eigen cation (\cf{H9O4+}),
serving as representative 
finite sized clusters, are compared in detail in this study 
to the hexagonal phase of ice~I in order to systematically 
study
hydrogen bonding as a function of temperature
from ambient conditions down to about 1~K.
Using these quite different system classes,
various hydrogen bond characteristics
and strengths 
can be studied in detail.
These range from 
rather long and thus weak 
hydrogen bonds in the neutral systems
to 
asymmetric double well situations in
the larger protonated water clusters
to a centered hydrogen bond in the Zundel cation
which is often called ``ultra-strong''. 
It is, therefore, expected that our results
can be transferred to various other hydrogen
bonded systems
depending on their hydrogen bond stengths.

\section{Computational details}
\label{sec:comp-det}

To study the temperature dependence of different hydrogen
bonded systems down to \SI{1.67}{\kelvin}, path integral molecular
dynamics simulations have been performed with the 
\texttt{i-PI}~program~\cite{Ceriotti2014/10.1016/j.cpc.2013.10.027}
and the \texttt{CP2k} program package~\cite{CP2K,Hutter2014/10.1002/wcms.1159}.
The potential energy surface of the cationic and thus
strong hydrogen bonded clusters, from
the Zundel cation (\cf{H5O2+}) 
over the protonated water trimer (\cf{H7O3+}),
to the Eigen cation (\cf{H9O4+}), was described
using a recently introduced neural network potential (NNP)
fitted to coupled cluster reference calculations~\cite{Schran2019/submitted}
that is used here to carry out molecular dynamics simulations via \texttt{CP2k}.
This NNP, 
which describes all investigated protonated species on equal footing, 
has been shown to 
not only
match the reference coupled cluster
theory with very high precision,
but is also able to accurately describe proton transfer in these protonated species.
The interactions in the three neutral systems of choice, 
i.e. 
the water dimer, water hexamer and hexagonal ice I$_\text{h}$,
were described 
by the q-TIP4P/F force field~\cite{Manolopoulos2009/10.1063/1.3167790},
which has been parametrized for the purpose
of path integral simulations of water, 
via calls to \texttt{LAMMPS}~\cite{Plimpton1995/10.1006/jcph.1995.1039}.
That water model,
where nuclear quantum effects effectively included in the parameterization of
the original TIP4P model have been explicitly eliminated,
has been shown to reproduce
many of the fascinating properties of liquid water and ice with
sufficient agreement to experiment~\cite{Manolopoulos2009/10.1063/1.3167790,
Ramirez2010/10.1063/1.3503764,
Herrero2011/10.1063/1.3559466,
Herrero2011/10.1103/PhysRevB.84.224112,
Habershon2011/10.1039/C1CP21520E,
Markland2012/10.1073/pnas.1203365109,
Kapil2018/10.1021/acs.jpcb.8b03896}.
In addition, it allows for fast and uncomplicated
exploration at different conditions, which makes it perfectly
suited in the present case.
Recall that 
this 
non-reactive 
water model does not
allow one to describe 
water dissociation and thus 
proton transfer by construction, which is 
an excellent approximation for our neutral
water systems,
whereas the reactive NNP is mandatory to simulate
the protonated water clusters being prone to 
proton transfer events.
Application of this force field
to the water hexamer was previously shown to
provide slightly inadequate
populations of individual isomers~\cite{Babin2013/10.1016/j.cplett.2013.06.041}.
This is expected to be of minor importance
in the present case, since
the individual properties of the hydrogen bonds
in the various isomers of the water hexamer are
very similar, which has been explicitly tested.
The systems of interest
were simulated at temperatures of
300, 
250, 100, 20, 10 and \SI{1.67}{\kelvin}
including the quantum nature of the nuclei
in conjunction with PIGLET
thermostatting~\cite{Ceriotti2012/10.1103/PhysRevLett.109.100604}
which has been recently extended to 
and validated at
ultra--low temperatures~\cite{Uhl2016/10.1063/1.4959602}.
In order to reach convergence, the path integral was
discretized using $P=$~6, 8, 16, 64, 128, and 256
replica at $T=$~300, 250, 100, 20, 10, and \SI{1.67}{\kelvin},
respectively.
The convergence of 
these path integral discretizations 
has been validated explicitly in Ref.~\citenum{Schran2018/10.1021/acs.jctc.8b00705}
for the prototypical hydrogen bond in the Zundel cation
and was chosen accordingly.
In case of the water hexamer, which features different
stable isomers, the interconversion was explicitly sampled
between 100 and \SI{300}{\kelvin},
while below \SI{100}{\kelvin}
the system remained in its starting configuration.
Therefore, different isomers were chosen as
the starting point below \SI{100}{\kelvin}.
In the following, results below \SI{100}{\kelvin} 
are exclusively shown for the water
hexamer in an ordered hexagonal ring
in order to allow for
a direct comparison with the hexagonal phase
of ice.
It is noted that the temperature dependence of the other
isomers below \SI{100}{\kelvin} is very similar to
the results presented in the following.
All reported simulations were propagated for at least
\SI{125}{\pico\second}
using a formal molecular dynamics
timestep of \SI{0.25}{\femto\second}
where \SI{2}{\pico\second} at the beginning of each
simulation were discarded as equilibration.

The condensed phase simulations were carried out 
using proton disordered 
ice~I$_\text{h}$ supercells
with periodic boundary conditions including 96 water molecules.
This setup has been adapted from the established 
initial conditions underlying 
Refs.~\citenum{Drechsel-Grau2014/10.1103/PhysRevLett.112.148302,Drechsel-Grau2014/10.1002/anie.201405989}.
It features two hydrogen ordered hexamer rings,
while the positions of the other hydrogen atoms were
chosen to minimize the box dipole moment according to the
usual ice rules.
The cell parameters (3$a$,2$\sqrt{\text{3}}$$a$,2$c$), where $a$ and $c$
are the standard hexagonal lattice parameters of ice I$_\text{h}$,
were chosen according to Ref.~\citenum{Roettger1994/10.1107/S0108768194004933,Roettger2012/10.1107/S0108768111046908} for each simulation
temperature.
It is noted that while this work has been in progress,
more precise measurements of the temperature dependent
lattice parameters of ice I$_\text{h}$ have become
available~\cite{Fortes2018/10.1107/S2052520618002159,Buckingham2018/10.1103/PhysRevLett.121.185505}.
However, for the purpose of this work we do
not expect large differences on the reported properties.

In order to validate the 
q-TIP4P/F force field
description of the interactions in ice at ultra--low temperatures,
\textit{ab initio} PIMD simulations were
performed at \SI{1.67}{\kelvin}.
These simulations were carried out 
with the \texttt{CP2k} program package~\cite{CP2K,Hutter2014/10.1002/wcms.1159}.
The electronic structure was solved on--the--fly 
via the \texttt{Quickstep} 
module~\cite{VandeVondele2005/10.1016/j.cpc.2004.12.014}
using the RPBE exchange-correlation
functional~\cite{Hammer1999/10.1103/PhysRevB.59.7413}
together with the D3~dispersion 
correction~\cite{Grimme2010/10.1063/1.3382344}
taking into account the two--body terms and applying zero damping. 
The charge density was represented on a grid up to a plane wave cutoff of 500~Ry.
The TZV2P basis set together with Goedecker--Teter--Hutter
pseudopotentials to replace the
core electrons 
of
the oxygen atoms~\cite{Goedecker1996/10.1103/PhysRevB.54.1703}
was used for the description of the Kohn--Sham orbitals.
The SCF cycles were converged to an error of $\epsilon_\text{SCF} = 10^{-7}~\text{Ha}$.
This electronic structure setup has been shown to describe
many properties of water in close agreement to
experiment~\cite{Morawietz2013/10.1021/jp401225b,
Forster-Tonigold2014/10.1063/1.4892400,
Morawietz2016/10.1073/pnas.1602375113}.
In total \SI{20}{\pico\second} 
(after an equilibration period of \SI{1.5}{\pico\second} 
using a starting configuration obtained from 
a well-equilibrated q-TIP4P/F force field simulation)
were generated 
for analyses
where the path integral has been discretized in terms of 48~replicas
in conjunction with the PIGLET algorithm~\cite{Ceriotti2012/10.1103/PhysRevLett.109.100604,Uhl2016/10.1063/1.4959602}

All reported properties were
evaluated for hydrogen bonded configurations,
where the standard hydrogen bond criterion
based on a donor--acceptor distance of 3.5~\AA{}
and a hydrogen bond angle $\angle_\text{HOO}$ of
30$^\circ$ has been applied~\cite{Luzar1996/10.1103/PhysRevLett.76.928}.
Other hydrogen bond criteria were explicitly tested
but resulted only in minor differences with respect
to the reported results in agreement with earlier 
systematic studies on this topic~\cite{Kumar2007/10.1063/1.2742385}.

In order to estimate the hydrogen bond strength
in the various studied systems, we calculated
the binding energy per water monomer
of each system for the optimized
minimum energy configurations.
As usual, the binding energy was
obtained as the difference between
the bound system and the sum of the
isolated components, where
for the protonated clusters the
system was dissected into one
hydronium core and the remaining
water molecules.
This binding energy was afterwards
normalized by the number of water
molecules present in the individual systems.
\section{Results and Discussion}
\label{sec:res}
\begin{figure}[tb]
    \centering{}
    \includegraphics[width=0.9\linewidth]{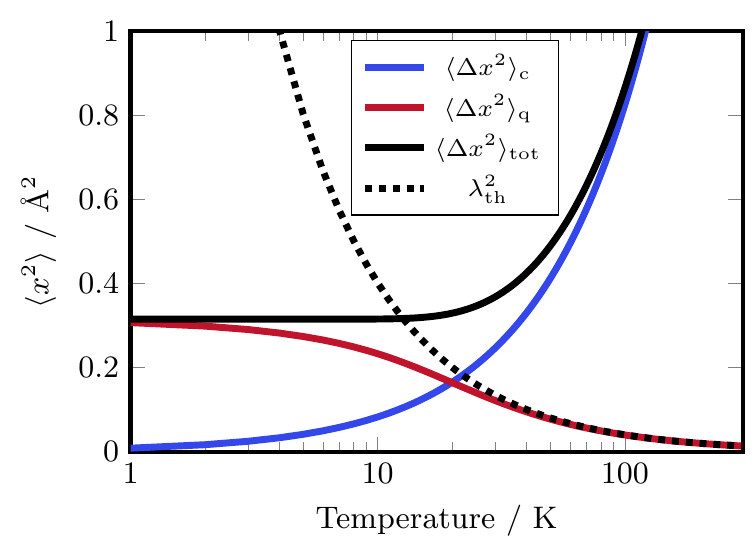}
    \caption{
        Comparison of the temperature dependent position fluctuations
        of a hydrogen atom trapped in a 
        1D--harmonic potential with a period
        of $2\pi\cdot$\SI{100}{\femto\second}.
        The total fluctuations $\langle \Delta x^2\rangle_\text{tot}$ 
        are shown in black,
        the classical fluctuations $\langle \Delta x^2\rangle_\text{c}$ in blue
        and the quantum fluctuations $\langle \Delta x^2\rangle_\text{q} = \langle \Delta x^2\rangle_\text{tot} - \langle \Delta x^2\rangle_\text{c}$ in red, 
where $\langle \cdots \rangle_\text{c}$ and $\langle \cdots \rangle_\text{tot}$
are the classical and quantum statistical Maxwell--Boltzmann (canonical) averages
respectively, at the given temperature. 
        The squared thermal wavelength $\lambda_\text{th}^{2}$ of a free
        hydrogen atom
        in one dimension
        is included for comparison as dotted line.
    }
    \label{fig:ho}
\end{figure}

In order to both, motivate 
and illustrate
the 
drastically changing relative 
importance of NQEs for hydrogen bonding 
when decreasing the temperature from ambient to
ultra--cold conditions,
we start by discussing the temperature
dependence of the total position fluctuations $\langle \Delta x^2\rangle_\text{tot}$
of a hydrogen atom trapped in a 1D--harmonic potential
serving as a simple textbook example.
These total fluctuations
can be straight--forwardly decomposed into the respective
quantum delocalization, $\langle \Delta x^2\rangle_\text{q}$,
and the purely classical position fluctuations $\langle \Delta x^2\rangle_\text{c}$
as shown in Fig.~\ref{fig:ho}.
The quantum contribution 
is the difference between a purely
classical description of the nuclei and
the correct total fluctuations
at the respective temperature.
For the simple
harmonic oscillator, 
the quantum delocalization $\langle \Delta x^2\rangle_\text{q}$
ultimately 
converges to a constant value when the temperature is lowered,
where it provides a measure of the 
harmonic 
zero point motion,
whereas it decays to strictly zero in the high temperature (classical) limit
as shown in Fig.~\ref{fig:ho}. 
In contrast, the classical (thermal) fluctuations $\langle \Delta x^2\rangle_\text{c}$
strictly vanish at zero temperature but increase linearly as a function of 
temperature, see Fig.~\ref{fig:ho}
(note the logarithmic temperature scale). 
The most interesting regime is at intermediate temperatures where 
both, quantum and thermal fluctuations contribute significantly, 
as indicated in Fig.~\ref{fig:ho} by the crossing 
of the $\langle \Delta x^2\rangle_\text{q}$ and $\langle \Delta x^2\rangle_\text{c}$ curves, 
which can be used to define the 
so-called ``crossover temperature'' 
between quantum and classical behavior~\cite{Weiss2012}.

Obviously,
realistic 
molecular 
systems that include hydrogen bonds are governed by 
pronounced 
anharmonic 
covalent and noncovalent 
interactions, which adds complexity to 
the undergraduate textbook 
picture 
that is illustrated with the help of Fig.~\ref{fig:ho}.
These anharmonic effects 
together with the pronounced many-body nature of the underlying intermolecular interactions
are the major reason for
the rich characteristics of the hydrogen bond,
ranging from 
weak and thus strongly asymmetric such non-covalent bonds 
with sizable proton transfer barriers 
to low-barrier situations 
to even centered 
or ultra-strong
hydrogen bonding as for example 
all 
encountered during the phase
transition from molecular ice~VIII to ionic
ice~X~\cite{Benoit2002/10.1103/PhysRevLett.89.145501}.
Therefore,
a systematic study is conducted 
in what follows 
in order to understand 
quantitatively
how hydrogen bonding is influenced
as a function of temperature down to the ultra--cold regime
depending on the hydrogen bonding class in the aforementioned sense. 
In experiment only the total quantum solution is
accessible, which is why we base our analysis purely
on quantum simulations.
For that purpose, strongly hydrogen bonded systems,
namely the 
protonated water dimer (Zundel complex), trimer and tetramer (Eigen complex) 
as well as three weaker hydrogen bonded systems,
being 
the water dimer and hexamer in vacuum and
the hexagonal phase of ice~I, ice I$_\text{h}$, 
have been selected to serve as representative cases. 
In these systems the gradual increase of hydrogen bonding and
resulting cooperative effects~\cite{Stokely2008/10.1073/pnas.0912756107,
Guevara2016/10.1039/c6cp00763e}
can be studied in detail,
while additionally 
granting
access to general differences and similarities between 
finite clusters and the condensed phase
as well as charged and neutral systems.
Note that in
our neutral
systems the 
rather long
hydrogen bonds 
lead to very 
high proton transfer barriers
and thus greatly suppress  proton transfer,
while 
the
protonated water clusters 
are much more prone to proton transfer events.
\begin{figure*}[t]
    \centering{}
    \includegraphics[width=0.9\linewidth]{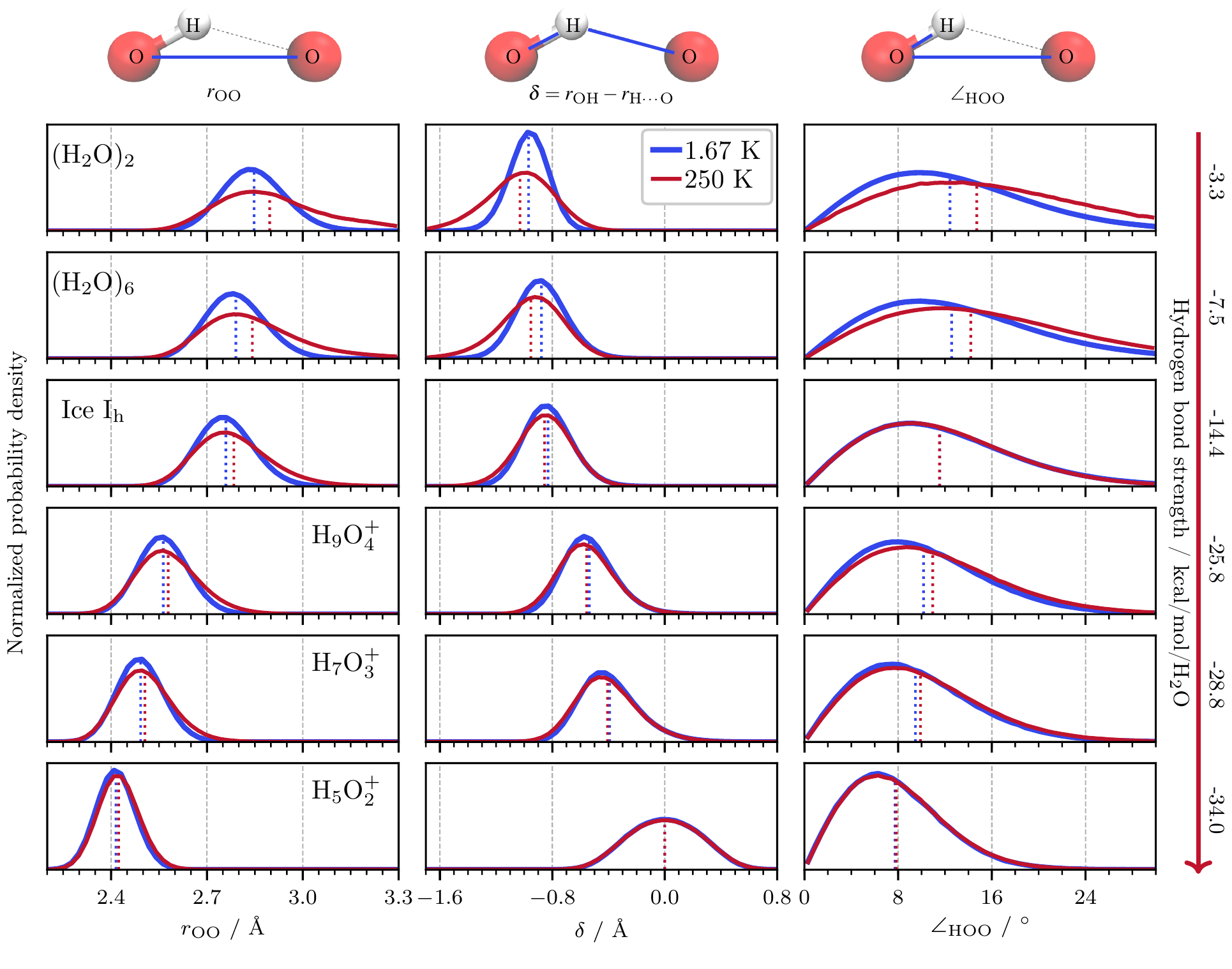}
    \caption{
        Normalized probability distributions of the heavy atom
donor--acceptor 
        distance $r_\text{OO}$ (left),
        the proton sharing coordinate $\delta$ (middle) defined
        as $\delta=r_\text{OH}-r_{\text{H}\cdots\text{O}}$,
        and the hydrogen bond angle $\angle_\text{HOO}$ (right)
        from PIMD simulations at 1.67~K (blue)
and 250~K (red) for 
the water dimer (\cf{(H2O)2}), water hexamer (\cf{(H2O)6}),
ice~I$_\text{h}$, Eigen cation (\cf{H9O4+}),
protonated water trimer (\cf{H7O3+}), and
Zundel cation (\cf{H5O2+})
from top to bottom. 
The different systems were ordered by increasing
hydrogen bond strength 
(cf. decreasing average $r_\text{OO}$ distance and $\angle_\text{HOO}$ angle)
from top to bottom
according to the estimated hydrogen bond strength
as described in Sec.~\ref{sec:comp-det}.
        The average of the individual distributions
        is marked with a vertical dotted line.
        Exclusively hydrogen bonded configurations are considered
        in this analysis based on the hydrogen
        bond criterion of Ref.~\citenum{Luzar1996/10.1103/PhysRevLett.76.928}.
    }
    \label{fig:ice_struc}
\end{figure*}
As illustrated 
based on the simple harmonic oscillator example 
the
relative 
importance of the quantum nature of the nuclei
increases when the temperature is lowered, but
for most structural properties, convergence with temperature
is
found to be 
reached relatively fast.
In order to investigate structural properties,
the distribution functions
of three main properties of the hydrogen bond
for the different systems are compared in Fig.~\ref{fig:ice_struc}
at 250 and \SI{1.67}{\kelvin}.
The chosen 
properties,
namely the 
donor--acceptor 
distance $r_\text{OO}$,
the proton sharing coordinate $\delta$, 
and the hydrogen bond angle 
$\angle_\text{HOO}$, are ideally suited to characterize
hydrogen bonding and 
all directly 
relate to the strength of the hydrogen bond. 
In general, shorter heavy atom distances, more symmetric
sharing coordinates, and smaller hydrogen bond angles
correlate well with stronger hydrogen 
bonding~\cite{%
Novak2007,
Schmidt2007/10.1021/jp074737n,
Michaelides2011/10.1073/pnas.1016653108,
Galkina2017/10.1134/s0022476617050080},
which 
allows
us to order the six systems from 
top to bottom in that figure.
To further quantify the hydrogen bond strength
in the studied systems, we evaluated
the binding energy per water molecule
as described in Sec.~\ref{sec:comp-det}.
The resulting 
rough
estimate of the hydrogen bond
strength is also included in Fig~\ref{fig:ice_struc}
(right vertical scale) 
and can be used to provide a comparison for
other hydrogen bonded systems, not included
in the present study,
in order to judge their dependence 
on nuclear quantum effects depending on the
temperature regime.

Let us first discuss in detail the heavy atom distance $r_\text{OO}$
at the two temperatures.
While for the water dimer at \SI{1.67}{\kelvin}
a relatively broad distribution with an average
at about 2.85~\AA{} is found, this distribution is 
shifted
towards smaller distances to an average
of $\approx$~2.80~\AA{} 
in the water hexamer and finally to
about 2.75~\AA{} in the bulk
where it is also much more narrow.
When moving to the stronger hydrogen bonds of
the protonated water clusters, the low
temperature distributions become even sharper
and are shifted to shorter $r_\text{OO}$ distances.
The distribution peaks at 2.56~\AA{} for the Eigen
cation, while being moved to 2.49~\AA{} for
the protonated water trimer and finally
to 2.41~\AA{} for the Zundel cation.
In case of the
neutral
finite clusters, increasing the temperature to
\SI{250}{\kelvin}
causes 
distinct changes,
more specifically broadening the distributions
and shifting them to 
larger distances.
In the condensed phase the temperature effect is
less dominant, while it diminishes further
when moving to the protonated water
clusters.
Note that the 
$r_\text{OO}$ distributions
of the Zundel cation are almost indistinguishable
although the temperature difference is about
\SI{250}{\kelvin}.
Secondly, 
the proton sharing coordinate $\delta$,
defined as the difference between the covalent
bond length and the 
noncovalent bond length (i.e. the hydrogen bond distance) 
$\delta = r_{\rm OH} - r_{\rm H\cdots O}$,
is analyzed to identify the degree
of asymmetry in the hydrogen bond.
This coordinate is systematically
shifted 
towards a more symmetric hydrogen bond
when comparing the water dimer, hexamer, ice,
and the three protonated water clusters.
This can be seen from the averages that
decrease from $-$0.95~\AA{} in the 
water dimer
to
$-$0.90~\AA{} in the hexamer to
$-$0.83~\AA{} in ice.
For the Eigen cation, the average of
the distribution is around $-$0.55~\AA{},
which is shifted to $-$0.40~\AA{} in
the protonated water trimer and finally
to zero for the Zundel cation,
indicating a 
centered
hydrogen bond.
Again, the distributions of the
neutral
finite sized
clusters are 
broader
and shifted to even more asymmetric
values
at \SI{250}{\kelvin}, while very similar
distributions are obtained
for ice
and the three protonated water clusters
at both temperatures.
As already found for the hydrogen bond length, the
distributions of the Zundel
cation are essentially indistinguishable
at both temperatures.

Finally, the hydrogen bond angle $\angle_\text{HOO}$
is selected to characterize the degree of deviation
from collinearity in the hydrogen bond,
which is the third proxy that correlates with the strength of hydrogen bonds. 
Here, the distributions of the
neutral
finite clusters 
peak at slightly larger angles than in the
condensed phase
and also feature a longer tail
towards larger angles, 
thus implying
weaker
hydrogen bonding in the
neutral
finite systems. 
In case of the protonated clusters, the
distributions are shifted further
to smaller angles when going
from the Eigen to the Zundel cation.
As before, temperature does 
essentially not affect the respective distribution in ice
and the protonated water clusters,
whereas it shifts it to
larger angles in the 
neutral finite clusters by about 2--3$^{\circ}$.
Thus, temperature weakens the hydrogen bond further
in the water dimer and hexamer compared to ice. 
Overall, these
results reveal clear trends
in hydrogen bonding 
of the three neutral systems
when the coordination
of the individual water molecules is
increased from 
finite sized clusters to
the condensed phase.
Due to the higher degree of hydrogen bonding
and the resulting cooperative
effects~\cite{Guevara2016/10.1039/c6cp00763e},
overall shorter,
more symmetric and more linear,
and 
therefore 
stronger hydrogen bonds are observed
in the condensed phase.
Interestingly, an opposite cooperative
effect is observed when analyzing the
protonated water clusters
(while keeping in mind their different hydrogen bonding topology). 
All structural properties
indicate for these systems 
and topologies
that the addition of further water molecules
is weakening the hydrogen bond,
since the donor--acceptor distance
is increased, the sharing coordinate
moves to more asymmetric values
and the hydrogen bond angle is
progressively 
increased. 
This can be explained by the larger degree
of charge delocalization when more
water molecules solvate the 
protonic defect.

In addition, this comparison of different structural
properties of the hydrogen bond reveals
that the temperature effects from 250 down
to \SI{1.67}{\kelvin} have 
only a minor
impact on the structural properties 
for the stronger hydrogen bonds,
or even no impact in case of the strongest (centered) hydrogen
bond as offered by the Zundel complex
(thus confirming pioneering work~\cite{Marx1997/10.1126/science.275.5301.817}). 
This reveals that for the Zundel cation
the main contribution to the structural properties of the
hydrogen bond comes from the nuclear zero point energy
even at a temperature as high as \SI{250}{\kelvin}.
Still, temperature has a bigger
impact on the larger protonated water clusters and ice, as seen
from the detailed comparison of the distributions
between \SI{1.67}{} and \SI{250}{\kelvin}.
Note that these observations are also in line
with a recent spectroscopic investigation of
the Eigen cation, for which only a very mild temperature
dependence of the spectrum has been
reported~\cite{Fagiani2016/10.1039/c6cp05217g}.
At the same time, the
neutral
finite clusters feature a systematic weakening
of hydrogen bonding when increasing the temperature.
When the results for the
different hydrogen bonded
systems are compared directly,
it can be concluded that 
minor temperature effects are observed
for strong hydrogen bonds
when moving from \SI{1.67}{} to \SI{250}{\kelvin},
but weaker hydrogen bonds feature
larger changes at ambient conditions compared to ground
state dominated temperatures.
This underlines that ultra--low temperature
conditions are ideally suited to study
strong hydrogen bonds, while
for weaker hydrogen bonds the results
need to be carefully transferred to ambient
conditions.
Let us next analyze the contribution of
NQEs 
on the quantum de/localization
of the individual atoms participating in the
hydrogen bond,
i.e. oxygens and hydrogens.
As introduced above for the harmonic oscillator,
this property features a strong dependence on the temperature.
Indeed, 
recent studies at ultra--low temperature have revealed an unexpected
localization of the nuclear wavefunction in finite hydrogen bonded
complexes~\cite{Walewski2013/10.1080/00268976.2013.822112},
especially for the light hydrogen atoms participating in intermolecular
interactions.
This effect is even further increased due to helium--solute
interactions~\cite{Walewski2014/10.1063/1.4870595}.
At the same time, the heavy 
donor and acceptor 
atoms of these hydrogen bonded clusters are less affected
and can even be more 
quantum 
delocalized than the much
lighter 
protons
in the hydrogen bond.
Recall that the
thermal de~Broglie wavelength $\lambda_\text{th}$
of a free particle scales with $\nicefrac{1}{\sqrt{M}}$.
From this consideration it is usually expected
that protons are much more quantum delocalized
than the heavier oxygen nuclei at the same temperature, which
highlights the unexpected character of 
the previously discovered~\cite{Walewski2013/10.1080/00268976.2013.822112,Walewski2014/10.1063/1.4870595}
larger delocalization of the oxygen atoms
in these hydrogen bonds at temperatures on the order of 1~K. 
This so--called ``interaction induced localization effect''
on protons trapped in hydrogen bonds
is far from being understood and needs further investigation
as a function of temperature.
\begin{figure*}[t]
    \centering{}
    \includegraphics[width=1.0\linewidth]{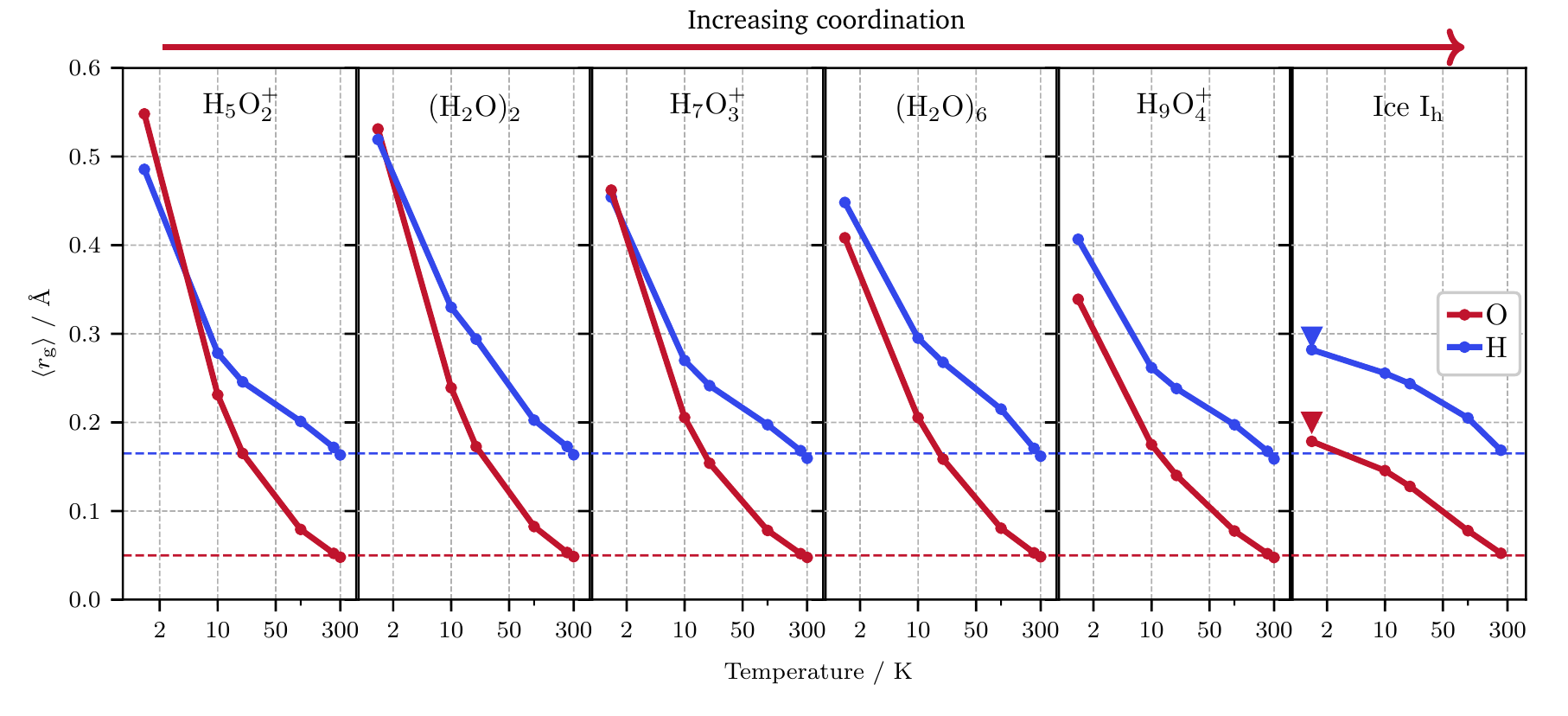}
    \caption{
        Temperature dependence of the quantum delocalization
        of the oxygen (red) and hydrogen atoms involved in hydrogen bonds (blue)
        measured by the averaged radius of gyration $\langle r_{\rm g}\rangle$
in 
the Zundel cation (\cf{H5O2+}), water dimer (\cf{(H2O)2}),
protonated water trimer (\cf{H7O3+}), water hexamer (\cf{(H2O)6}),
Eigen cation (\cf{H9O4+}), and ice~I$_\text{h}$
from left to right. 
The different systems were ordered by increasing
coordination from left to right.
In order to highlight the similarities
between the different systems, the
average values of all systems at
the highest considered
temperatures
are included as horizontal dashed lines
for the oxygen and hydrogen atoms, respectively.
        Exclusively hydrogen bonded configurations are considered
        in this analysis.
        All results
for the neutral systems
        are obtained using the q-TIP4P/F water force
        field,
while the cationic systems were described by a
highly accurate neural network potential 
(NNP) of coupled cluster accuracy (see text). 
        For ice I$_\text{h}$, the results at the lowest
        temperature were validated using 
ab initio path integral simulations based on RPBE--D3 (see text) 
        electronic structure 
        (triangles).
    }
    \label{fig:deloc}
\end{figure*}
In order to systematically investigate such
effects
for the various systems of the present study
the quantum delocalization of the
oxygen atoms as well as the hydrogen atoms
participating in hydrogen bonds are analyzed in detail.
Quantum delocalization 
of nuclei at a given temperature
can be conveniently separated from
the total fluctuations in the path integral formalism
by measuring the instantaneous spread of the path
integral ring polymer
with reference to its centroid position. 
The resulting averaged radius of gyration of the discretized path integral
\begin{align}
r_{\mathrm g}^2=\frac{1}{P}\sum_{s=1}^P\left<\left(\textbf{R}_s-
\textbf{R}_{\mathrm c}\right)^2 \right>
, 
\end{align}
is depicted as a function of temperature in Fig.~\ref{fig:deloc} 
for the individual atoms.
This property, if squared, is identical to the quantum
fluctuations $\langle \Delta x^2\rangle_\text{q}$ that were
introduced above.
In addition, in the case of a free particle, 
the radius of gyration is equivalent to the thermal de~Broglie
wavelength and thus gives access to the quantum
delocalization of 
the interacting nuclei of interest
at a certain temperature.
At 
\SI{300}{\kelvin}
the oxygen atoms feature
only a relatively small average radius of gyration
in all
systems
according to Fig.~\ref{fig:deloc}.
This indicates that their quantum 
delocalization adds  
a rather small contribution 
to the classical description of the oxygen atoms.
At the same temperature, the hydrogen atoms in
all three systems are considerably more 
quantum
delocalized.
This 
not only
highlights the importance of NQEs on
the properties of the hydrogen bond already
close to ambient conditions,
but also confirms general expectations
based on the stark differences of the thermal de~Broglie
wavelength of heavy versus light free particles. 
Together, these data support the well--known picture 
that heavy atoms, such as oxygen nuclei, behave
essentially like classical point particles
at ambient temperature, while 
the light protons are already affected by NQEs
and thus ``smeared out'', which is 
independent of the particular system. 
Most interestingly,
all chosen systems
covering a broad range of 
hydrogen bond strengths
feature essentially the same radius of gyration
for the hydrogen and oxygen nuclei
as seen by the respective horizontal dashed lines in Fig.~\ref{fig:deloc}.
As the temperature decreases, both
the oxygen and hydrogen atoms in 
all 
systems become less localized,
but a relatively constant
offset between the hydrogen and oxygen atoms is
initially 
conserved.
However, for 
the three smallest finite clusters
this offset starts decreasing
at about \SI{10}{\kelvin} to finally
result in a slightly larger average radius
of gyration of the oxygen
nuclei compared to the protons
at around
\SI{1}{\kelvin}.
This effect is 
most pronounced
for the centered
(or ``ultra-strong'')
hydrogen bond present in the Zundel cation, \cf{H5O2+}.
In other words: The quantum delocalization of
the heavy oxygen nuclei exceeds that of the 
light protons in this temperature regime
as opposed to the usual scenario that is
recovered at higher temperatures. 
Exactly this ``interaction induced localization'' of the
protons in hydrogen bonds at ultra--low temperatures
has first been described in 
Ref.~\citenum{Walewski2013/10.1080/00268976.2013.822112}
for various hydrogen bonded dimers~--
including even the heavier chlorine nucleus
when HCl is involved in hydrogen bonding. 
It is important to note that this effect is not an 
artifact
of the underlying 
force field
description of the interactions
as used previously~\cite{Walewski2013/10.1080/00268976.2013.822112}.
Indeed, very similar results are obtained for
the Zundel cation, \cf{H5O2+},
and the protonated water trimer, \cf{H7O3+},
on an essentially
converged potential energy surface fitted to
coupled cluster reference calculations~\cite{Schran2019/submitted}.
The water hexamer
and the Eigen cation reveal
a similar temperature
dependence, but the oxygen atoms remain
more localized than the hydrogen atoms
at the lowest considered temperature, although
the difference between the two atoms 
is significantly descreased at 1~K 
compared to \SI{250}{\kelvin};
it is tempting to speculate that the crossover
might set in at still lower temperatures.
In stark contrast to the finite sized systems, the 
difference in the quantum delocalization 
between oxygen and hydrogen atoms 
as provided by the radius of gyration
remains almost constant
over the whole temperature range for 
ice I$_\text{h}$ as seen in the right panel of Fig.~\ref{fig:deloc}.
Note that also these 
force field
results were validated by
\textit{ab initio} PIMD simulations
using the RPBE--D3 density functional 
(see Sec.~\ref{sec:comp-det} for the details), 
providing 
a very similar
delocalization at the
lowest considered temperature 
(see the
triangles
in the right panel of Fig.~\ref{fig:deloc}).
In addition, the extent of the quantum delocalization
of both atoms is overall 
significantly
smaller compared to 
the finite clusters
at the lower temperatures and,
at variance with the behavior found for these finite clusters,
seems to flattens out. 
These results reveal that the
``interaction induced localization effect''
found for finite clusters in vacuum can not
be transferred to 
ice. 
At the same time, almost the same quantum
delocalization is found at higher temperatures
for all 
systems, which clearly
needs to be 
analyzed
in more detail.

We now 
focus on an explanation for
the different localization effects of
the hydrogen bond in isolated clusters 
versus 
hexagonal ice. 
Key to the understanding of the differences
is the coordination of the individual hydrogen
bond partners in the various systems.
As indicated by the ordering in Fig.~\ref{fig:deloc}
this coordination gradually increases when
moving to the larger systems.
While the two dimers feature only a single
hydrogen bond, every water molecule
in the 
(cyclic)
water hexamer participates
in two hydrogen bonds, and
the hydronium core of the Eigen
cation forms three hydrogen bonds.
Finally, in ice I$_\text{h}$ every
water molecule donates and accepts
two hydrogen bonds, resulting in
a coordination 
number of four per molecule. 
The increase in the coordination
clearly correlates with the
overall size of the atoms
at ultra--low temperatures
and additionally with the
crossover between oxygen and hydrogen atoms.
As illustrated above using the simple harmonic
oscillator model,
the constraining potential of an atom
shows the biggest impact at low
temperatures, while
at higher temperatures
essentially the thermal wavelength of the 
respective free
particles is 
recovered. 
This explains why the coordination 
modulates the 
size of the atoms mainly at low temperatures,
while essentially the same radii of gyration are
observed at ambient conditions.
It can therefore be concluded that 
enhanced
coordination of the atoms, which 
counteracts
translational and rotational delocalization
and constraints the 
quantum fluctuations of the
atoms, modulates the
``interaction induced localization effect''
and entirely prevents it in the condensed phase.

Overall, the detailed analysis
of quantum delocalization in the hydrogen
bond reveals that the previously reported
``interaction induced localization effect``
can be found in finite sized clusters, 
whereas it is absent in the condensed phase. 
In addition, this difference is traced back to
the coordination
of the involved atoms.
Given the systematic study of various systems,
it is expected that these results
are rather general for comparing
finite sized clusters and condensed phase
hydrogen bonded systems.

\section{Conclusions}
\label{sec:conclusion}
In summary, 
hydrogen bonding has been investigated
in detail
as a function of temperature,
from ambient (300~K) down to ultra-cold ($\approx 1$~K)
conditions,
including nuclear quantum effects
for various hydrogen bonded systems,
in particular
for the water dimer and hexamer
as well as for protonated water clusters from the Zundel to the Eigen complex,
and 
hexagonal ice~I$_\text{h}$.
Our analysis revealed
minor temperature effects on
structural properties of the hydrogen bond,
specifically its length, angle and a/symmetry,
for strongly hydrogen bonded systems,
such as protonated water clusters 
including the Zundel complex, 
when moving from about 1~K up to ambient
conditions.
In stark contrast, weaker 
hydrogen bonds, as present in the investigated neutral
systems
and notably in the water dimer, 
were found to be significantly
altered when moving from ground state dominated
temperatures to ambient conditions, where
mainly additional weakening of this noncovalent bond
is observed.
These systematic results
unveil that
ultra--low temperature experiments
such as tagging photodissociation and 
helium nanodroplet isolation spectroscopy
are ideally suited to study strong hydrogen bonds,
while for weaker hydrogen bonds the results need to be carefully
transferred to ambient conditions.
Furthermore, the quantum delocalization of 
the nuclei involved in 
hydrogen bonding
was explored in detail to reveal
considerable differences between finite clusters
in vacuum and the condensed phase.
In contrast to the condensed phase,
the quantum delocalization of the heavier oxygen atoms
was found to be comparable to, or even larger than that
of the hydrogen atoms in systems with relatively weak
spatial constraints.
These differences were traced back to the 
coordination 
of the involved atoms which
counteracts
translational and rotational quantum delocalization.
In addition, 
the coordination was
shown to be the main driving force behind
the intriguing larger localization of
the lighter hydrogen atoms compared to
the oxygen atoms as previously reported for
hydrogen bonded clusters 
in superfluid helium nanodroplets
based on quantum simulations in explicit bosonic helium. 

\section*{Acknowledgements}
It gives us great pleasure to thank Harald Forbert, Fabien Brieuc, and
Felix Uhl for helpful discussions.
This research is part of the Cluster of Excellence 
``RESOLV''
(EXC~2033, ID~390677874) funded by the \textit{Deutsche Forschungsgemeinschaft}, DFG.
C.S. acknowledges partial financial support from the 
\textit{Studien\-stiftung des Deutschen Volkes} as well as from the
\textit{Verband der Chemischen Industrie}.
The computational resources were provided by HPC@ZEMOS,
HPC-RESOLV, and BOVILAB@RUB. 

\end{document}